\documentstyle[12pt,aaspp4,epsf,flushrt]{article}

\begin{document}

\title{Heating of Intergalactic Gas and Cluster Scaling Relations}
\author{Michael Loewenstein\altaffilmark{1}}
\affil{Laboratory for
High Energy Astrophysics, NASA/GSFC, Code 662, Greenbelt, MD 20771}
\altaffiltext{1}{Also with the University of Maryland Department of Astronomy}

%===============================================================================
\begin{abstract}

X-ray observations of galaxy groups and clusters are inconsistent with
the predictions of the simplest hierarchical clustering models,
wherein non-baryonic and baryonic components are assembled together
under the sole influence of gravity.  These departures are in the
sense that the intergalactic medium is hotter and more extended than
expected, and become increasingly strong for less massive systems. I
model these effects by constructing baseline sequences of hydrostatic
polytropic models normalized to observations of high-temperature
clusters and numerical simulations, and then transforming them by
adding proscribed amounts of heat per particle at the cluster
center. I present sequences with a universal value of this heating
parameter that simultaneously reproduce recently published observed
(gas and total gravitational) mass-temperature and entropy-temperature
relations. The required amount of energy injection is consistent with
constraints on the number of supernovae needed to account for observed
intracluster silicon abundances, provided that energy injection is
centrally concentrated. I argue that most of the heating occurred
during or after the assembly of the cluster, and not exclusively in
pre-collapse proto-cluster fragments.

\end{abstract}


%===============================================================================
\section{Context}

Hierarchical clustering in a universe where non-baryonic dominates
over baryonic matter is the primary heuristic framework for studying
large scale structure formation; and, one of the primary laboratories
for disconfirming variations deriving from particular combinations of
fluctuation spectrum and cosmological world model is the ensemble of
clusters of galaxies.\footnote{For simplicity -- and because there is
no universally accepted demarcation -- I follow Jones et al. (1998) in
referring to all bound collections of galaxies as clusters, therein
subsuming the usual definition of groups of galaxies.}  Such
distinguishing characteristics of the cluster population include the
two-point correlation function and mass function of collapsed
non-baryonic halos (e.g., \cite{g99}).

Since non-baryonic matter is not amenable to observation, these
relationships must be studied indirectly through gravitational lensing
and baryonic tracers such as stars (in galaxies) and hot intergalactic
gas.  The latter, perhaps the most unambiguous tracer because of the
density-squared dependence of the hot gas emissivity, is the focus of
the present study. The now well-developed database of cluster X-ray
observations made throughout this decade by the {\it ROSAT} and {\it
ASCA} observatories have provided accurate measurements of the cluster
temperature function (e.g., \cite{m98}), and log N -- log S relation
(e.g., \cite{j98}); however, mapping from these to the relevant
non-baryonic halo relations requires knowledge of the thermal and
dynamical structure and history of the intracluster medium (ICM).

In the most well-studied hierarchical clustering scenarios, structure
formation is merger-driven, proceeding sequentially from small to
large scales. A remarkable result of numerical simulations along these
lines is that the mass density profiles of relaxed, non-baryonic halos
can be characterized by a universal functional form (\cite{nfw} --
hereafter NFW; \cite{j99}) -- at least over some large radial range.
These halos are not precisely self-similar, since less massive halos
tend to collapse earlier when the universe was more dense and thus
have -- on average -- more concentrated non-baryonic density
distributions. Subject to collisional processes such as shocks, the
structure of the baryonic component will differ from that of the
non-baryonic, but in a regular and predictable manner if the two
components simply approach equilibrium in concert (e.g., \cite{enf}).

However, there is ample observational evidence for departures from
such simplicity -- evidence that becomes more striking as one focuses
on less massive systems. The most well-established such breakdown is
the relative steepness of the correlation between X-ray luminosity and
ICM temperature ($T_x$), with $L_x\propto {T_x}^{\sim 3}$ observed in
contrast to $L_x\propto {T_x}^{\sim 2}$ predicted if gravitational
collapse is the sole mechanism driving the evolution of both
components (e.g., \cite{a99}). Likewise, values of $T_x$ very near the
non-baryonic component virial temperature are predicted, an equality
that evidently increasingly breaks down for less massive systems based
on departures from $\mu m_p\sigma^2=kT_x$ (e.g., \cite{bmm}) and
$M\propto {T_x}^{3/2}$ (\cite{hms}; \cite {e99} -- hereafter EF),
where $\mu m_p$ is the mean mass per particle, $\sigma$ the
one-dimensional cluster velocity dispersion, and $M$ the total mass
(evaluated at $r_{500}$: the radius within which the average density
contrast, relative to the critical density, is 500). There is also
evidence for a concomitant increase in ICM extent from the steepness
of the $M_{ICM}$-$T_x$ correlation (where $M_{ICM}$ is the ICM gas
mass within $r_{500}$; \cite{mme}), and the gradual flattening towards
low temperatures of the relation between entropy and $T_x$ (the
``entropy floor;'' \cite{pcn} -- hereafter PCN).

All of these deviations from predictions where ICM structure is
exclusively determined by gravitational collapse are indicative of an
important role for extra-gravitational heating in determining ICM
properties and their variation with mass. Understanding the magnitude
and source of the heating not only is necessary for mapping from
observed X-ray properties to those of the non-baryonic component, but
also sheds light on galaxy formation since feedback from the earliest
star formation epoch is the most likely heating mechanism.  Since most
metals in the universe are produced in objects of high overdensity
(\cite{c99}), and are subsequently ejected into intergalactic space by
supernova-driven galactic outflows (\cite{p99}), measurements of ICM
heavy metal enrichment provide independent constraints on the
magnitude of energy injection (\cite{w91}, \cite{l96}).

I address this set of interconnected puzzles in the present work as
follows. Adopting the most widely-used functional form for the
non-baryonic density distribution that is completely determined by a
single concentration parameter, baseline mass sequences of
equilibrium, polytropic ICM configurations are constructed.  The
sequences are anchored at the high mass end using the results of
high-temperature ICM observations and simulations (that are mutually
consistent), and then extended to lower masses following specified
mass-scalings of the non-baryonic halo concentration.  A thermodynamic
perspective proves physically illuminating (\cite{b97}; \cite{bbp}),
and so heated families of mass sequences are calculated that differ
from the baseline sequences by fixed amounts of heating at the cluster
center. It then remains to be seen whether the mass sequence in a
particular family can simultaneously explain all of the observed
deviations from the scaling relations predicted by models without
heating, while also conforming with the nucleosynthetic constraints
from ICM abundance studies. I will show that, in fact, they do so.

%===============================================================================
\section{The Baseline Configuration}

In this section I construct simple analytic models that accurately
reproduce the observed and simulated ICM distribution of the most
massive clusters. This serves as a useful point of departure for
creating heated distributions for the entire range of cluster masses.

\subsection{Polytropic Models}

I consider spherically symmetric, hydrostatic gas configurations in a
total gravitational potential following the NFW prescription. Thus,
\begin{equation}
{1\over {\rho_{\rm gas}}}{{dP_{\rm gas}}\over {dr}}=
-{{GM_{\rm grav}(<r)}\over {r^2}},
\end{equation}
where $\rho_{\rm gas}$ and $P_{\rm gas}$ are the ICM density and
pressure distributions, respectively.  $M_{\rm grav}$ is the total
gravitational mass given by
\begin{equation}
M_{\rm grav}(<r)=M_{\rm virial}{{F_{\rm NFW}(r/a_{\rm NFW})}\over {F_{\rm NFW}(c)}},
\end{equation}
where 
\begin{equation}
F_{\rm NFW}(x)\equiv{\rm ln}(1+x)-{x\over {1+x}};
\end{equation}
$a_{\rm NFW}$ is the non-baryonic mass distribution scale-length, $c$
the concentration parameter $r_{\rm virial}/a_{\rm NFW}$, and $r_{\rm
virial}$ the virial radius.  Models that are analytic and conveniently
flexible are obtained under the polytropic assumption, $P_{\rm
gas}\propto {\rho_{\rm gas}}^{1+1/n}$ (\cite{ssm}; \cite{cmt}; EF).
The polytropic gas distribution is determined by the single function
\begin{equation}
\theta(x)\equiv 1+A\left[{{{\rm ln}(1+x)}\over x}-1\right].
\end{equation}
The gas temperature and density follow from $T_{\rm
gas}=T_o\theta(r/a_{\rm NFW})$ and $\rho_{\rm
gas}=\rho_o\theta^n(r/a_{\rm NFW})$, respectively, where $T_o$ and
$\rho_o$ are the (assumed finite) central values of gas temperature
and density. The constant
\begin{equation}
A\equiv {{2c}\over {F_{\rm NFW}(c)}}{{T_{\rm virial}}\over {(n+1)T_o}}
\end{equation}
must not exceed a maximum value to assure non-negative temperature and
density out to $r_{\rm virial}$, {\it i.e.} there is a minimum ratio
of $T_o/T_{\rm virial}$ for each pair $(c,n)$.  In the isothermal
limit ($n\rightarrow \infty$), the density distribution is given
instead by
\begin{equation}
\rho_{\rm gas}=\rho_{\infty}\theta_{\rm iso}(r/a_{\rm NFW}),
\end{equation}
where
\begin{equation}
\theta_{\rm iso}(x)=(1+x)^{A_{\rm iso}/x},
\end{equation}
and 
\begin{equation}
A_{\rm iso}\equiv {{2c}\over {F_{\rm NFW}(c)}}{{T_{\rm virial}}\over {T_{\rm gas}}}
\end{equation}
(\cite{mss}; EF). The gas density at the origin for the isothermal
case is related to that at infinity by $\rho_o=\rho_{\infty}e^{A_{\rm
iso}}$.

The virial mass, radius, and temperature are connected through the
relationships
\begin{equation}
M_{\rm virial}={{4\pi}\over 3}{r_{\rm virial}}^3\Delta_c\rho_{\rm crit}
\end{equation}
and
\begin{equation}
{{kT_{\rm virial}}\over {\mu m_p}}
={1\over 2}{{GM_{\rm virial}}\over {r_{\rm virial}}},
\end{equation}
where $\Delta_c$ is the density contrast (with respect to the critical
density $\rho_{\rm crit}=9.21\ 10^{-30}{h_{70}}^2$ g cm$^{-3}$;
$h_{70}$ is the Hubble constant in units of 70 km s$^{-1}$ Mpc$^{-1}$)
within which virial equilibrium is established.

In addition to the concentration parameter, $c$, and polytropic index,
$n$, I parameterize models by quantities more observationally relevant
than $T_o$ and $\rho_o$ -- the global gas fraction, $f_{\rm
virial}\equiv f_{\rm gas}(r_{\rm virial})$, and
density-squared-weighted average temperature in virial units,
$\tau$. It follows that
\begin{equation}
{{T_o}\over{T_{\rm virial}}}=\tau{{I_{2n}(c)}\over {I_{2n+1}(c)}},
\end{equation}
and 
\begin{equation}
{{\rho_o}\over{\rho_{\rm crit}}}={{\Delta_c}\over 3}c^3f_{\rm virial}
I_{n}(c)^{-1},
\end{equation}
where
\begin{equation}
I_m(c)\equiv{{\int_0}^c}dx\ x^2\theta(x)^m.
\end{equation}
Equation (11) must be solved iteratively.

\subsection{The Baseline Model}

In order to establish a baseline sequence of models without
extra-gravitational heating, I first turn to the literature on
numerical simulations. In particular, I consider the average $z=0$
model in an $\Omega_o=0.3$, $\Lambda_o=0.7$, $h_{70}=1$ cosmology from
Eke et al. (1998). Although the demographic properties of the cluster
population greatly depend on world model, the $z=0$ structure of the
most massive clusters is relatively insensitive to this choice. The
non-baryonic component in the simulated cluster is completely
characterized by $M_{\rm virial}=1.51\ 10^{15}{h_{70}}^{-1}$
M$_{\odot}$, $\Delta_c=100$ ($r_{\rm virial}=2.99{h_{70}}^{-1}$ Mpc,
$kT_{\rm virial}=6.9$ keV), and $c=6.47$; the gas component by $f_{\rm
virial}=0.087$ (by assumption), $T_x\approx T_{\rm virial}$
($\tau=1$), and
\begin{equation}
\rho_{\rm gas}\approx\rho_{\rm crit}\delta_o
\left(1+{{r^2}\over {{r_{\rm core}}^2}}\right)^{-{{3\beta_{\rm fit}}/2}},
\end{equation}
with $\delta_o=1970$, $r_{\rm core}/r_{\rm virial}\approx 0.05$, and
$\beta_{\rm fit}=0.735$.

Temperature profiles measured with {\it ASCA} have been characterized
by $n\approx 5$ polytropic distributions (\cite{mfs}); however, both
temperature (\cite{ibe}) and density profiles in hydrostatic models
with this index are significantly steeper than those in the
simulations.  In fact, the density distribution in the ICM of the
simulated cluster is accurately recovered only in models approaching
the isothermal limit. Moreover, the best fit has $\tau=1.1$ rather
than $\tau=1$. Figure 1 shows a comparison between the simulated
cluster density profile expressed by equation (14) and models with
$f_{\rm virial}=0.087$ and $(n,\tau)=(5,1.1), (100,1),$ and
$(100,1.1)$ -- clearly the last provides the superior fit.  The cause
of the deviation of the temperature profile in the numerical
simulation, with its mild gradient at large radii, from its
characterization in the baseline model is evidently due to residual
unthermalized bulk kinetic energy in the simulated ICM at $r>0.5
r_{\rm virial}$.  This has little effect on the observed cluster-wide
temperature -- generally, an emission-weighted average over $\sim 0.5
r_{\rm virial}$ -- that is the the global temperature used in the
correlations considered in this paper.

In a recent paper, EF describe the results of their analysis of {\it
ROSAT} PSPC data on the ICM in 36 high-temperature clusters.  The mean
parameters that emerge from surface brightness fitting, using
functions of the forms expressed in both equations (6) and (14),
display a remarkable agreement with the above. They find mean values
of $\beta_{\rm fit}$ and $a_{\rm NFW}/r_{\rm core}$ of 0.72 and 3.17
compared to 0.735 and 3.08 in the simulations of Eke et
al. (1998). However, while (on average) $A_{\rm iso}=11.3$ in the
latter, EF find $A_{\rm iso}=10.3$ from fitting the surface brightness
profile using equation (6) -- corresponding to $\tau=1.1$ as expected
(Figure 1). The mass-temperature relationship derived by EF is also
consistent with $\tau=1.1$ at the high-$T_x$ end.

The appropriate value of the global gas fraction, $f_{\rm virial}$,
can be estimated from the values of $M_{\rm ICM}$ at the highest $T_x$
(where effects of heating should be minimized) calculated within
$r_{500}$ by Mohr et al. (1999).\footnote{ICM masses derived in EF
(1999) generally agree with those in Mohr et al. (1999) for clusters
in common to the two papers, although there are significant
discrepancies in gas mass fraction due to differences in adopted
temperatures and mass estimation technique.}  Extrapolating to $r_{\rm
virial}$ using an $n=100$ polytrope with $\tau=1.1$ in an NFW
potential with $r_{\rm virial}/a_{\rm NFW}=6.2$ corresponding to
$kT_x=10$ keV ($M_{\rm virial}=2.3\ 10^{15}{h_{70}}^{-1}$ M$_{\odot}$)
yields $f_{\rm virial}=0.155{h_{70}}^{-3/2}$ (extrapolated from
$f_{\rm gas}=0.136{h_{70}}^{-3/2}$ at $r_{500}$).

To summarize, I adopt as baseline configurations hydrostatic models
with a polytropic index of $n=100$ normalized such that the
emission-measure-weighted average temperature is $1.1\times$ the
virial temperature, and gas fraction within the virial radius $f_{\rm
virial}=0.155{h_{70}}^{-3/2}$ (${\rho_o}=5050{h_{70}}^{-3/2}{\rho_{\rm
crit}}$). The gravitating mass is assumed to follow equations (2) and
(3), with
\begin{equation}
c=c_{10}\left({M_{\rm virial}\over {M_{10}}}\right)^{-\gamma},
\end{equation}
where $M_{10}=2.3\ 10^{15}{h_{70}}^{-1}{\rm M}_{\odot}$ corresponds to
the virial mass at a density contrast of 100 for a $kT_x=10$ keV
cluster, and $c_{10}=6.2$ is adopted from Eke et al. (1998) assuming
$\gamma=0.1$ (appropriate for a CDM-like fluctuation spectrum; NFW).
These models yield extremely accurate representations of the density
distributions of observed average high-temperature clusters (as well
as of numerical simulations of the same), and do so for the correct
observed mean temperature. Any proposed heating mechanism must
therefore have relatively little effect on the ICM of the most massive
clusters.

The adopted models are virtually isothermal. Models where the inverse
of the polytropic index is substantially different than zero (or where
the temperature distribution is otherwise strongly varying with
radius) have density distributions that are too steep to be consistent
with the observations (as well as the simulations) if the ICM is in
hydrostatic equilibrium in an NFW potential (this remains true for
other concentration parameters than those considered above).  Thus, if
the temperature profiles reported in Markevitch et al. (1998)
(recently brought into question by Irwin et al. 1999 and \cite{f99})
hold up, either hydrostatic equilibrium, electron-ion equipartition
(\cite{e97}; \cite{t98}), or the assumption of a NFW-type total mass
profile must break down well within the virial radius.

%===============================================================================

\section{Heated Families of Polytropic Mass Sequences}

I now describe a simple prescription for constructing families of mass
sequences for polytropic intracluster media in hydrostatic equilibrium
in an NFW potential that deviate from the baseline sequence described
above by having been heated by differing amounts.  Heated families are
completely determined by specifying the final gas fraction and
polytropic index for each cluster mass. I show, for several different
families, that a mass sequence corresponding to the heating needed to
explain the observed entropy floor also reproduces the other
departures from gravity-only self-similarity described in Section 1.

\subsection{Formalism}

Consider a transformation from the baseline polytropic configuration
to a new one:
\begin{equation}
\{n^{bl}, {T_o^{bl}}, {\rho_o^{bl}}\}\longrightarrow\{n, T_o, \rho_o\},
\end{equation}
where the $bl$ superscript refers to the baseline model parameters.
The changes in specific entropy and heat corresponding to the
transformation are
\begin{equation}
\Delta s=\int ds  = {3\over 2}{k\over {\mu m_p}}
\left[{\rm ln}{{T_o}\over {T_o^{bl}}}-
{2\over 3}{\rm ln}{{\rho_o}\over {\rho_o^{bl}}}\right],
\end{equation}
and
\begin{equation}
\Delta q=\int Tds  = {3\over 2}{k\over {\mu m_p}}\left(T_o-T_o^{bl}\right)
\left[1-{2\over 3}{1\over {T_o-T_o^{bl}}}
\int {{d{\rm ln}\rho}\over {d{\rm ln}T}}dT\right],
\end{equation}
where $ds=(3k/2\mu m_p)d{\rm ln}(T/\rho^{2/3})$ and the integrals
extend from the baseline to the new configuration.  One can define an
effective adiabatic index, $\Gamma$, such that
\begin{equation}
{{T_o}\over {T_o^{bl}}}=\left({{\rho_o}\over {\rho_o^{bl}}}\right)^{\Gamma-1},
\end{equation}
from which it follows that
\begin{equation}
\Delta s=C_{\Gamma}^{-1}{k\over {\mu m_p}}{\rm ln}{{T_o}\over {T_o^{bl}}}
\end{equation}
and
\begin{equation}
\Delta q=C_{\Gamma}^{-1}D_{\Gamma}{k\over {\mu m_p}}\left(T_o-T_o^{bl}\right)
=D_{\Gamma}{{T_o-T_o^{bl}}\over {\rm ln}({{T_o}/{T_o^{bl}}}})\Delta s,
\end{equation}
where 
\begin{equation}
C_{\Gamma}\equiv {{2\Gamma-2}\over {3\Gamma-5}},
\end{equation}
and 
\begin{equation}
D_{\Gamma}=1+C_{\Gamma}\left(\Gamma-1\right)^{-1}
\left(1-{{\Gamma-1}\over {T_o-T_o^{bl}}}
\int {{d{\rm ln}\rho}\over {d{\rm ln}T}}dT\right).
\end{equation}

For a constant heating adiabat
\begin{equation}
{{d{\rm ln}\rho}\over {d{\rm ln}T}}={1\over {\Gamma-1}}; D_{\Gamma}=1;
\end{equation}
and, therefore
\begin{equation}
T_o={T_o^{bl}}+C_{\Gamma}T_q; \rho_o=
{\rho_o^{bl}}\left(1+C_{\Gamma}{{T_q}\over {T^{bl}}}\right)^
{1\over{\Gamma-1}},
\end{equation}
where $T_q$ is defined such that, at the center, the new configuration
has an extra heat per particle, $\Delta q=kT_q/\mu m_p$ relative to
the baseline configuration. Non-decreasing entropy requires $\Gamma\le
1$ (temperature increasing as density decreases) or $\Gamma\ge 5/3$
(temperature increasing as density increases).  $\Gamma=1$,
$\Gamma=0$, and $\Gamma=\infty$ correspond to isothermal, isobaric,
and isochoric transformations, respectively; for $\Gamma=1$ equation
(25) is replaced by
\begin{equation}
T_o=T_o^{bl}; \rho_o={\rho_o^{bl}}e^{-{{T_q}\over {T^{bl}}}}.
\end{equation}

I assume that the amount of heating per particle is the same for all
clusters (and therefore less effective for more massive clusters), so
that
\begin{equation}
\tau_q\equiv {{T_q}\over {T_{\rm virial}}}
=\epsilon\left({{M_{\rm virial}}\over {M_{10}}}\right)^{-{2\over 3}},
\end{equation}
where $\epsilon$ is the (assumed universal) injected heat per particle
in units of 10 keV.  To make a clearer connection with observationally
relevant quantities, in place of $T_q$ and $\Gamma$, $\epsilon$ and
the slope $\nu$ are specified, assuming that the global gas fraction
in the heated configuration is given by
\begin{equation}
f_{\rm virial}=f_{10}\left({{M_{\rm virial}}\over {M_{10}}}\right)^\nu,
\end{equation}
where $f_{10}$ is the global gas fraction for a cluster with a 10 keV
ICM ($f_{\rm virial}$ is generally assumed constant with mass along
baseline mass sequences).

The prescription for calculating heated families of polytropic mass
sequences can be summarized as follows.  (1) Baseline mass sequences
are determined from specification of the concentration-mass relation
(the parameter $\gamma$ in equation 15), $f_{10}$
($0.155{h_{70}}^{-3/2}$), $n^{bl}$ (100), and $\tau$ (1.1). (2) A
heated family is then computed, having specified $\nu$ and $n$, by
varying $\epsilon$.  In practice, for each mass (and its appropriate
$f_{\rm virial}$ from equation 28 and concentration $c$ from equation
15) hydrostatic models are iterated on the parameter $\Delta\tau\equiv
(T_o-{{T_o}^{bl}})/T_{\rm virial}$ until the desired value of
$\epsilon$ is achieved.  The relationship between $\epsilon$ and
$\Delta\tau$ is approximated as
\begin{equation}
\epsilon={3\over 2}\left({{M_{\rm virial}}\over {M_{10}}}\right)^{2\over 3}
\Delta\tau\left(1-{2\over 3}{{{\rm ln} (\rho_o/{\rho_o^{bl}})}\over
{{\rm ln}(1+\Delta\tau/\tau)}}\right),
\end{equation}
which is exact if the heating is along an adiabat with a constant
index $\Gamma$.

%===============================================================================

\subsection{A Simple Heated Family}

For simplicity, I initially consider a heated family where the gas
fraction, $f_{\rm virial}$, and polytropic index, $n$, are identical
to those in the baseline model ({\it i.e.}  $\nu=0$ and $n=100$; also
$\gamma=0.1$).  Perhaps the most striking evidence for significant
heating of the ICM in a form similar to that of equation (27) is the
entropy floor at low $T_x$ recently reported by PCN, where the
``entropy'' is defined as $kT_x{n_e}^{-2/3}$; $n_e$ is the electron
density in cm$^{-3}$ evaluated at $r_c$ ($r_c$ is defined as
one-tenth the radius within which the total density contrast is 200,
as estimated from $T_x$ using the baseline sequence of models), and
$kT_x$ is in keV.  The value of $\epsilon$ that produces an
entropy-temperature relation with the observed entropy floor at
$\approx 100$ keV cm$^2$ is 0.35.\footnote{This value is sufficiently
large that is has a non-negligible effect on even the hotter clusters,
and I have reduced the dimensionless average temperature $\tau$ in the
baseline model from 1.1 to 1.05 to preserve the high level of
agreement with the observed density profiles for high-$T_x$ clusters.}
This (dashed) curve is shown in Figure 2 along with observed data
points (assuming $h_{70}=1$) from analysis of {\it ASCA} imaging data
(\cite{f97}; \cite{h99}).  Also shown is the value from numerical
simulations (filled triangle; Eke et al. 1998), and the (solid) curve
for $\epsilon=0$. Note that the latter is slightly steeper than linear
due to the more concentrated non-baryonic mass distributions for
cooler clusters.

Remarkably, this same simple heated mass sequence is in excellent
agreement with all of the other observed trends alluded to in Section
1.  Figure 3 show a plot of total mass evaluated at $r_{500}$ versus
$T_x$ for $\epsilon=0$ (solid curve) and $\epsilon=0.35$ (dashed
curve).  While most of the heating in the $\epsilon=0.35$ sequence
goes into decreasing the gas concentration, rather than increasing the
temperature, less massive clusters are predicted to become
increasingly hotter than expected in the absence of heating. For this
sequence, $\Gamma\approx 0.8$ and $C_{\Gamma}\approx 0.15$ (see
equations 19 and 22), nearly independent of mass, so that all clusters
have an increase in internal energy per particle of
$C_{\Gamma}kT_q\approx 0.5$ keV.  The resulting predicted
mass-temperature relationship is
\begin{equation}
M(r_{500})\approx 1.32\ 10^{15}
\left({{kT_x-0.5\rm keV}\over {9.5\rm keV}}\right)^{3/2}
Q(M){h_{70}}^{-1} {\rm M}_{\odot}.
\end{equation}
$Q(M)$ is a form factor relating $M(r_{500})$ to $M_{\rm virial}$ and
increases from 1 at $kT_x=10$ keV to 1.12 at $kT_x=1$ keV for
$\gamma=0.1$ in equation (15).  Equation (30) is in excellent accord
with estimates from X-ray observations compiled by Horner et
al. (1999; data points in Figure 3) and the best-fit mass-temperature
trend from EF (stars in Figure 3).  A similar accordance is found for
the temperature dependence of gas mass within $r_{500}$ (Figure 4,
where the stars represent the best-fit trend from Mohr et
al. 1999). In a similar vein to Figure 3, Figure 5 shows the
$\epsilon=0$ and $0.35$ model variations of $(kT_{\rm virial}/\mu
m_p)^{1/2}$ with $T_x$ that can be characterized by
\begin{equation}
kT_{\rm virial}\approx {{kT_x-0.5\rm keV}\over {1.05}},
\end{equation}
compared to observed values of the velocity dispersion $\sigma$ from
Girardi et al. (1998; data points) and the best-fit $\sigma$-$T_x$
correlation of Bird et al. (1995; stars). The agreement with the
best-fit trend is excellent; although, a large observed scatter is
evident, and I have not considered velocity dispersion anisotropies
and gradients that are possible sources of discrepancy between
$kT_{\rm virial}$ and $\mu m_p\sigma^2$.

To summarize, if one considers cluster non-baryonic matter and gas
distributions as predicted by large scale structure numerical
simulations, and then transforms the gas distribution by adding 3.5
keV per particle of heat at the center while preserving the global
baryon fraction and polytropic index, the observed trends wherein the
ICM becomes increasingly hotter and more extended\footnote{Mohr et
al. (1999) claim an absence of evidence of a more extended ICM in
cooler clusters because of the lack of a clear trend of the average
gas particle location with temperature; however, the scenario
described above predicts that this does not become noticeable until
temperatures below those of the clusters in their sample.}  are
simultaneously and accurately reproduced.

%===============================================================================

\subsection{Other Heated Families}

With families of mass sequences anchored at high-$T_x$ as described in
Section 2.2, there remains the freedom to vary the following three
(one for the gravitating mass, two for the ICM) slope parameters: the
mass-scalings of total matter concentration -- $\gamma\equiv-{\rm
dln}c/{\rm dln}M_{\rm virial}$ -- and gas fraction within the virial
radius -- $\nu\equiv{\rm dln}f_{\rm virial}/{\rm dln}M_{\rm virial}$,
and the polytropic index in the heated configuration -- $n\equiv{\rm
dln}\rho_{\rm gas}/{\rm dln}T_{\rm gas}$. The implications of varying
these are as follows.

Since an increase in the concentration of the gravitating mass
distribution increases the central ICM density in the equilibrium
configuration, a steeper dependence of $c$ on $M_{\rm virial}$ (as
would be expected if the fluctuation spectrum is flatter than in CDM
cosmogonies; NFW) would imply that more heating (larger $\epsilon$) be
required to explain the observed entropy floor.  For example $\sim
4.5$ keV per particle would be required if $\gamma$ were increased
from 0.1 to 0.2 (that is the concentration increases by $\sim 2.3$
instead of $\sim 1.5$ for 1 keV, relative to 10 keV, clusters).

Positive values of the parameter $\nu$ correspond to cooler, less
massive clusters having progressively smaller baryon fractions within
the virial radius. The heating required to produce the observed
entropy floor can be reduced to $\sim 2.5$ keV per particle for
$\nu=0.25$ (implying a baryon fraction in 1 keV clusters nearly three
times lower than in 10 keV clusters). It is implicitly assumed that
the reduction in $f_{\rm virial}$ (by mass leaving the cluster, or
simply expanding beyond $r_{\rm virial}$) is a result of the heating:
lower values of $\epsilon$ can be accommodated if, instead, cooler
clusters are assumed to be intrinsically gas-poor. However, the
magnitude of the entropy gap between observed cool clusters and that
expected assuming no heating and constant $f_{\rm virial}$ would
require an extreme gas-fraction trend (PCN) of which there is no
observational evidence (EF, Mohr et al. 1999).

Finally, I have considered a family of mass sequences where the
polytropic index in the heated model decreases by 50\% for each factor
of two decrease in mass ({\it i.e.}  from $n=100$ to $n\approx 11$ as
$T_x$ -- defined as the emission-measure-averaged ICM temperature
within $r_{500}$ -- varies from 10 to 1 keV), as might be appropriate
if heating leads to more-nearly isentropic configurations (Balogh et
al. 1999). Since as $n$ decreases the ICM density distribution
steepens, larger values of $\epsilon$ -- corresponding to $\sim 4.5$
keV per particle -- are demanded by the observed entropy floor.

One can conclude from the above that the observed entropy floor for
cooler clusters is consistent with a mass sequence of polytropic
equilibrium configurations heated by $>3$ keV per particle at the
cluster center (see Figure 2; 3 keV is about the temperature at which
the observations start to deviate significantly from the predictions
of models without heating). In the context of these models, this same
amount of heat simultaneously accounts for the observed steepening of
the relationship between ICM mass within $r_{500}$ and temperature
(Figure 4).  Note that the former is a local indicator of ICM
expansion, while the latter is a global indicator. Also note that any
contrivance meant to lower the required amount of energy injection
(e.g., by reducing the intrinsic gas fractions of cooler clusters),
will destroy the simultaneous match to the observed steepening of the
relationship between total mass within $r_{500}$ and temperature
(Figure 3) or, equivalently, the increasing departure from $T_x=T_{\rm
virial}$ implied by the observed $\sigma$-$T_x$ relation (Figure 5).

\section{Discussion}

Given the striking success of the $\epsilon\sim 0.35$ heated family
mass sequence in reproducing the observed trends in global ICM
properties -- in particular their monotonically increasing departure
with decreasing temperature from expectations of models without
extra-gravitational heating -- it is worthwhile to consider the
implications for the relative gas distributions in hot and cool
clusters, the connection between heating and metal enrichment, and the
astrophysical interpretation of these models in the context of
hierarchical formation of clusters.

\subsection{Relative Gas Distributions}

Clearly, the existence of an entropy floor caused by heating implies
that cooler clusters will have more extended ICM, {\it i.e} both lower
central densities and shallower density slopes. And indeed this was
discovered by PCN in their analysis of {\it ROSAT} cluster data.  The
effect of heating on the gas distribution is perhaps most evident in
plots of the cumulative gas fraction. Figure 6 shows a comparison of
these for 1 and 10 keV clusters with and without heating (the latter
from the simple heated family of Section 3.2). In the absence of
heating, cooler clusters are expected to have significantly higher gas
fractions at small radii (relative to the virial radius) because of
their more concentrated mass distributions. Including heating, the
opposite is very strongly the case: at $r_{500}$ ($\sim 0.5r_{\rm
virial}$), 1 keV clusters are about twice as gas-poor as 10 keV
clusters (Figure 7), even though it is assumed that they have
identical gas fractions at $r_{\rm virial}$. A similarly strong trend
is seen for the ICM density distribution slope, as shown by Figure 8
which plots $\beta_{\rm image}\equiv 3{\rm dln}\rho_{\rm gas}/{\rm
dln}r$ at $r_{500}$ versus $T_x$ for mass sequences with and without
heating.  The density distribution is predicted to become flatter at
all radii greater than $0.1r_{\rm virial}$ for cooler clusters.

Whether these trends have been detected -- and, if so, to what
quantitative extent -- is currently problematic. Since both the
density slope and gas fraction vary (possibly non-monotonically) with
radius, comparisons must be made at radii corresponding to identical
density contrasts (e.g., $r_{500}$). Thus the total mass must be
estimated (for cooler systems this must done from the data itself, not
from an extrapolation of the hot-cluster mass-temperature relation) in
order to make a meaningful comparison of density slopes as well as gas
fractions.  Moreover, deriving the density distribution for the cooler
clusters ({\it i.e.}, groups) is greatly complicated by uncertainties
in background subtraction, in treating the emission from individual
galaxies, and possibly by departures from spherical symmetry
(\cite{z98}, \cite{h99}). Analysis along the lines of EF (who restrict
themselves to the more luminous, and therefore hotter, systems) on the
PCN ensemble of surface brightness profiles could prove illuminating.
Interestingly, extended elliptical galaxy X-ray halos, with $kT_x\sim
0.7$-1 keV, tend to have $\beta_{\rm image}\sim 0.5$, in accord with
the heating sequence of Figure 8 and their likely physical continuity
with ICM (\cite{b98}).

\subsection{The Connection Between Heating and Metal Enrichment}

The presently most reliable diagnostic of supernova heating in
clusters -- that presumably occurred during star formation driven
galactic outflows at early epochs -- is the ICM silicon
abundance. Only iron abundances are determined more accurately;
however, the amount of energy ejection associated with the mass of Fe
in the ICM is sensitive to the poorly determined (e.g., \cite{glm})
relative fractions of Fe originating from Type Ia (SNIa) and Type II
(SNII) supernovae.  Not only have accurate Si abundances been
determined for a large sample of clusters (\cite{f98}), but the
explosion energy per Si yield is similar in standard SNIa and SNII
models (e.g., Gibson et al. 1997) so that the energy injection
associated with the observed amount of Si is robust to uncertainties
in the SNIa/SNII ratio.  The equivalent temperature of the injected
energy can be expressed as
\begin{equation}
k{T_q}^{SN}\sim 1.6Z_{Si}e_{SN}\ {\rm keV},
\end{equation}
where $Z_{Si}$ is the ICM Si abundance, relative to solar, and
$e_{SN}$ is the average injected energy per supernova (perhaps boosted
by stellar winds from SNII progenitors; \cite{l95}) in units of
$10^{51}$ erg.

Since $Z_{Si}$ typically is $\sim 0.7$ for rich clusters (\cite{f98}),
supernova energy injection would seem to fall short by a factor of
$\sim 3$ of providing the required heating (this problem is
exacerbated for groups where the Si abundance is often estimated to be
lower by an additional factor of two; \cite{dmm}; \cite{h99}).
However, one must keep in mind that the requirement of 3 keV per
particle of heating is for the cluster center only.  If this heating
were uniformly present cluster-wide, the resulting expansion of the
ICM would lead to an extremely steep decrease of $f_{\rm virial}$ with
mass. The fact that this is not observed implies that the energy
injection decreases outward from the center. The central concentration
of SN energy injection may be reflected in abundance gradients that
have now been measured with {\it ASCA} (e.g., \cite{efm}).  Additional
heating at the cluster center could result from early AGN activity,
although the mechanism for efficiently dissipating AGN energy flux in
the ICM is unclear.

The average cluster-wide heating per particle can be estimated as
$5C_{\Gamma}kT_q/2\mu m_p\sim 0.38kT_q/\mu m_p$, where $C_{\Gamma}$ is
defined in equation (22) and $\Gamma\approx 0.8$.  Thus the
global-average heating is on the order 1 keV per particle. This is
more in line with that expected from nucleosynthetic considerations,
but requires a high energy deposition efficiency into the ICM or
$e_{SN}>>1$. The latter is possible if the initial mass function of
SNII progenitors is skewed toward high masses (\cite{tww}). This
amount of energy injection exceeds the escape temperature of the
lowest mass systems, and could lead to removal of some fraction of the
ICM -- depending on the detailed history of the heating with respect
to the dynamical evolution of the cluster.  As discussed in the
previous subsection, it is unclear whether lower mass systems have
systematically lower gas fractions.

\subsection{Physical Interpretation and the Epoch of Heating}

The amount of heat characterizing the $\epsilon\sim 0.35$ heated
family mass sequence corresponds to that required to transform the ICM
from a non-heated equilibrium state, and is thus an upper limit since
the gas may be pre-heated. If the heating occurs at sufficiently low
density ({\it i.e.}, prior to cluster collapse), the energetic
requirements for obtaining the observed entropy minimum are greatly
reduced.

PCN argue that this must be the case. However, this is based in part
on consideration of a SN rate (for the particular case of the Coma
Cluster) derived from the present stellar content and the assumption
of a standard initial mass function (IMF) that is well-established to
fall far short of providing the observed heavy metal enrichment (e.g.,
\cite{l96}).  Using standard SN yields, the Si enrichment, in solar
units, can be expressed as
\begin{equation}
Z_{Si}=0.2\left({{10M_{\rm stars}}\over {M_{\rm ICM}}}\right)
{{\eta_{SN}}\over {0.01}},
\end{equation}
where $M_{\rm stars}$ and $M_{\rm ICM}$ are the total cluster stellar
and ICM masses, respectively, and $\eta_{SN}$ is the total number of
SN per solar mass of present-day stars.  PCN, assuming
$\eta_{SN}=0.01$ as is appropriate for a standard IMF, thus derive an
equivalent supernova temperature $\sim 5$ times lower than the value
in equation (32) that is based on the actual observed rich cluster Si
abundance.  The relative epochs of cluster assembly and ICM heating
cannot be constrained solely on energetic grounds.

In pre-heating models the ICM is heated to some initial temperature
$kT_*$ and high entropy state.  The ICM in the lowest mass systems is
spared any further entropy increase via accretion shocks but, instead,
is accreted and compressed adiabatically -- thus preserving the
initial entropy at all radii.  An entropy floor of the kind reported
by PCN is predicted at low ICM temperature, and can be expressed as
\begin{equation}
n_e^{-2/3}kT=100\Sigma_o\ {\rm keV}\ {\rm cm}^2
\end{equation}
with $\Sigma_o\sim 1$.

Consider two extreme variations on the pre-heating scenario. In the
first (e.g., PCN), the universe as a whole is pre-heated to $kT_*$ at
a redshift of $z_*$ and, therefore, an electron density of
\begin{equation}
n_{e*}=\Omega_{\rm baryon}{{\rho_{\rm crit,o}}\over {\mu_em_p}}(1+z_*)^3
=1.2\ 10^{-7}(1+z_*)^3\ {\rm cm}^{-3}
\end{equation}
(\cite{t99}), where $\rho_{\rm crit,o}$ is the critical density at
zero redshift, and $\Omega_{\rm baryon}=0.0255{h_{70}}^{-2}$ as
implied by standard Big Bang nucleosynthesis. That is,
\begin{equation}
kT_*=2.45\ 10^{-3}\Sigma_o(1+z_*)^2\ {\rm keV}.
\end{equation}
Reproducing the observed luminosity- and mass-temperature relations
requires $kT_*\approx 0.5$ keV (Cavaliere et al. 1999), {\it i. e.}
$z_*\approx 13$ for $\Sigma_o=1$. In fact, however, to obtain the
change in slope of the $L_x$-$T_x$ relation (that is the transition to
adiabatic compression) at the correct temperature requires
$\Sigma_o=2.5$-10 (\cite{t99}), implying $z_*=3.5$-8. That is,
$\Sigma_o=1$ can be reconciled with the mass-temperature relationship
only if the universe was heated to 0.5 keV at a redshift greater than
10 -- and, in this case, the transition from shock heating to
adiabatic compression occurs at too low a mass to be consistent with
the $L_x$-$T_x$ relation. The transition occurs at the appropriate
mass if the level of the entropy floor has been underestimated by a
significant factor, however pre-heating redshifts greater than 3 are
still required if the mass-temperature relation is to be recovered as
well. Moreover, such a high initial entropy is predicted to lead to
baryon fractions in groups only $\sim 0.2$ of the universal value
(Cavaliere et al. 1999), which could be problematic since there are
group baryon fractions observed to exceed 10\% (\cite{m96},
\cite{h99}).

In the scenario of Balogh et al., each proto-cluster is heated at
turnaround; {\it i.e.} they assume
\begin{equation}
n_{e*}=\Omega_{\rm baryon}{{\rho_{ta}(z_*)}\over {\mu_em_p}}
\approx 8.3\ 10^{-6}g(z_*)(1+z_*)^3\ {\rm cm}^{-3}
\end{equation}
for $\Omega_o=0.3$ and $h_{70}=1$, where $g(z_*)=1$, 0.88, 0.81, and
0.75 at $z_*=1$, 2, 3, and 5, respectively. Thus,
\begin{equation}
kT_*\approx 1.2\ 10^{-2}g(z_*)^{2/3}\Sigma_o(1+z_*)^2\ {\rm keV};
\end{equation}
so that $kT_*=0.5$ keV implies $z_*=6.2$. The models of Balogh et
al. (1999) also require $\Sigma_o>1$ ($\Sigma_o\approx 3.7$) to
explain the low-temperature steepening of the $L_x$-$T_x$ relation.

Both pre-heating scenarios require an initial entropy several times
greater than the observed entropy floor as estimated by PCN.  These
can perhaps be reconciled if one recalls that the PCN entropy floor is
defined using a global average temperature.  Pre-heating implies that
accretion shocks become weak and the ICM increasingly isentropic at
low temperature (well below where the entropy- and
luminosity-temperature relations change slope, at $kT_x \sim 3$
keV). In the limit of pure adiabatic compression, significant
temperature gradients are expected. As a result the local entropy at
$r_c$ may be larger than the PCN estimate.

One can consider pre-heated mass sequences of models as defined in
Section 2, by fixing the polytropic index $n$ at 3/2. Such models are
normalized by fixing two of the following three parameters:
$\Sigma_o$, the mass-averaged temperature in units of the virial
temperature $\tau_m$, and the gas fraction within the virial radius
$f_{\rm virial}$. A general feature of these models is that the ICM
temperature at $r_c$ is about 50\% greater than the emission-averaged
temperature, and the entropy at $r_c$ correspondingly higher than that
estimated using the emission-weighted average.  Results for an
isentropic mass sequence with $\Sigma_o=2$ (chosen to approximately
reproduce the PCN entropy floor) and $\tau_m=1.05$ are shown by the
dotted curves in Figures 3 and 4 (sequences with constant $f_{\rm
virial}$ have similar properties at low $T_x$).  The mass-temperature
relation parallels that of the unheated sequence, shifted over by
$\sim 2$ in temperature, so that the observed relation is well-matched
at low $T_x$ (Figure 3).  The ICM mass-temperature relation is
reproduced as well for $kT_x<1.5$ keV (Figure 4), and only starts to
deviate strongly beyond 2 keV (since the gas fraction exceeds 0.5)
where the isentropic assumption should start to break down.

The temperature profile in the isentropic models are relatively steep
(Figure 9). This can be compared to the observed gas distribution for
HCG 62 -- the best measured group temperature profile in terms of
accuracy and extent.  The density profile is flat as predicted,
$n_e\sim r^{-1}$ (\cite{p93}); but, for an isentropic distribution the
temperature is expected fall to below 0.5 keV beyond 500 kpc -- and
does not (\cite{fpo}).

An assessment of the pre-heating scenario can now be summarized.  If
heating occurs prior to collapse, clusters with temperatures
sufficiently below the turnover of the luminosity- (or, equivalently,
the ICM mass-) and entropy-temperature relations are expected to be
isentropic and display a steep trend of decreasing baryon fraction
with decreasing temperature. The present observational evidence does
not support this, although the existing data is not of sufficient
quality to be conclusive. Moreover, to effect these turnovers at the
appropriate mass scale requires an initial entropy at least twice the
level of the observed entropy floor.  In order to simultaneously
explain the turnover in the mass-temperature relation the
proto-cluster gas must be pre-heated to $kT_*>0.3$ keV at redshift
\begin{equation}
z_*>11{\Sigma_o}^{-1/2}\left({{\mu_em_pn_{e*}}\over 
{\Omega_{\rm baryon}\rho_{\rm crit}(z_*)}}\right)^{1/3}-1,
\end{equation}
where the pre-collapse overdensity ($\mu_em_pn_{e*})/(\Omega_{\rm
baryon}\rho_{\rm crit}(z_*)$) must -- by definition -- lie between one
and the overdensity at turnaround ($\sim 15$). In the context of the
thermal history of baryons in the universe (\cite{o99}), such a level
of pre-heating seems unlikely at that high a redshift. In fact the
(mass-averaged) entropy in the baryon content of the universe is a
strongly decreasing function of redshift, the universe having been
both much cooler and more dense in the past, and does not approach the
level of the observed cluster entropy floor until relatively recently.

Thus, I would conclude that pre-heating models in their present
preliminary form cannot simultaneously explain all of the
heating-induced departures from self-similarity observed in clusters
-- although a definitive assessment must await more refined
pre-heating models and their comparison with more extensive and
complete observations of low-mass systems. On the other hand, the
success of the heated families of equilibrium polytropes in
simultaneously explaining the deviations from self-similarity observed
in clusters (Figures 2-5) suggests that substantial heating --
certainly in the cluster core -- occurred after the cluster was, at
least partially, in place. Perhaps some of the heating was triggered
by merging activity associated with the latter stages of cluster
formation. It must be noted, however, that the metal enrichment (or,
at least, the Fe enrichment) that presumably accompanies the heating
cannot have occurred much more recently than $z=1$ based on the lack
of evolution of cluster abundances (\cite{m97}), and the properties of
elliptical galaxies -- the sites of primordial star formation
generally presumed responsible for ICM heating and metal enrichment --
as a function of redshift (\cite{b99}, and references therein).

A detailed consideration of these issues requires a fully
time-dependent approach (e.g., \cite{k99}), and must reconcile the
required amount of heating with the observed properties of cluster
galaxies (Wu et al., in preparation) -- which may be problematic
(\cite{m99}).

\subsection{Cool Cluster Caveats}

The breaking, by energy injection, of gravity-only self-similarity by
an increasing amount for cooler clusters raises the following
caveats. (1) In deriving the cluster mass function from X-ray
luminosity ($L_x$) or temperature functions, care must be taken to
avoid using single slope $L_x$-$T_x$ or $M$-$T_x$ relations extending
all the way down from high $T_x$.  (2) The gas fraction is a fairly
steep function of radius for cool clusters (Figure 6) and values
measured out to, e.g, $r_{500}$, may not accurately reflect the global
value. Groups may appear gas-poor in comparison to rich clusters
while, in fact, their ICM are simply more extended with respect to
their non-baryonic components.

If hierarchical clustering correctly describes the formation of large
scale structure, cooler clusters will tend to form over a wider range
of redshifts (Balogh et al. 1999); and presumably, there will be
greater variation in the relative timing of merging/collapse and star
formation ({\it i.e.} cluster construction and heating). The large
scatter observed in metal abundance and gas fraction (e.g., Davis et
al. 1999) may be a reflection of this diversity.

\section{Summary and Conclusions}

In this paper I have attempted to construct a simple framework for
understanding the relationship between the baryonic and non-baryonic
components in galaxy clusters as manifest in the observed scaling of
total mass, and ICM mass and central entropy with ICM temperature
$T_x$.

Observed, as well as simulated, ICM density distributions of the most
massive clusters are well-characterized by nearly isothermal
hydrostatic configurations, with $f_{\rm virial}=0.155{h_{70}}^{-3/2}$
and $T_x=1.1T_{\rm virial}$, in an NFW potential. However, there are
clear and systematic departures from the expectations derived from
simply extending application of these baseline models to cooler, less
massive systems.  These departures are in the sense that the ICM in
less massive systems have an increasingly large excess of thermal
energy (from the steepening of the total mass-temperature relation),
and become increasingly less concentrated -- both globally (from the
steepening of the total mass-temperature relation) and, more
prominently, locally in the cluster core (from the flattening of the
central entropy-temperature relation).

This implies that substantial heating has occurred, heating with an
increasingly profound effect on less massive systems. I considered
transformations of the baseline models to families of new
configurations that have specified amounts of additional heating at
the cluster center (and possibly different gas fractions and
polytropic indices), and found that mass sequences heated by a
universal amount per particle can simultaneously reproduce the
observed total mass-temperature, ICM mass-temperature, and central
entropy-temperature trends to excellent precision. The required
central heat input is substantial, $>3$ keV per particle; however, the
cluster-wide average value is $\sim 1$ keV per particle: just
consistent with robust global nucleosynthetic constraints derived from
rich cluster Si abundance measurements and standard supernova
energies.  The concentration of the required heat injection is
reflected in the fact that the local (central) entropy-temperature
relation shows a more extreme departure from the gravity-only
prediction than the global ICM mass-temperature relation.

The 3 keV per particle of heating required at the centers of clusters
is an upper limit since it is derived assuming no pre-heating.  This
limit is derived from static considerations without recourse to a full
dynamical treatment (with its many accompanying parameters), and
becomes increasingly accurate the later the epoch of heating with
respect to cluster virialization.  PCN argue that the heating must
occur prior to the epoch of cluster assembly. However, this is based
in part on an underestimate of the available energy from supernova
explosions by a factor of 3-5.  If the universe as a whole is
pre-heated, the total mass-temperature relation implies heating the
baryons to $\sim 0.5$ keV at $z>3$ which is unlikely given that few
objects with that virial temperature will have collapsed and most star
formation is yet to occur.  Pre-heating becomes more feasible the
higher the overdensity of the proto-cluster at that epoch; however,
pure pre-heating models generally require a higher entropy floor than
is observed in order to reproduce the luminosity-temperature relation
(equivalent to the ICM mass-temperature relation successfully
reproduced by the heated mass sequence presented in Section 3.2). A
strong prediction of pre-heating models is that the lowest mass
clusters should be isentropic, a prediction difficult to assess with
presently existing data although the entropy in the $\sim 1$ keV group
HCG 62 appears to increase with radius.

The form of the observed departures from self-similarity would seem to
imply the majority of ICM heating, {\it i.e} much of the massive star
formation, was contemporaneous with -- and perhaps caused by -- the
assembly of the cluster, occurring at redshifts between 10 and 1.
Comparison of more detailed evolutionary models that incorporate time-
and space-dependent heating spread out over this epoch with improved
X-ray observations -- particularly of low-mass systems -- will greatly
illuminate the early star formation history of galaxies and the effect
this has on the surrounding environment.

\acknowledgments

I am grateful to U. Hwang for computing and providing the data points
and errorbars in Figure 2, to D. Horner for providing the data points
in Figures 3 and 5, and to A. Fabian, R. Mushotzky, P. Tozzi, and the
anonymous referee for providing useful feedback.

\clearpage

%===============================================================================

\clearpage

%===============================================================================
 
\begin{figure}
\centerline{\epsfxsize=0.8\textwidth\epsfbox{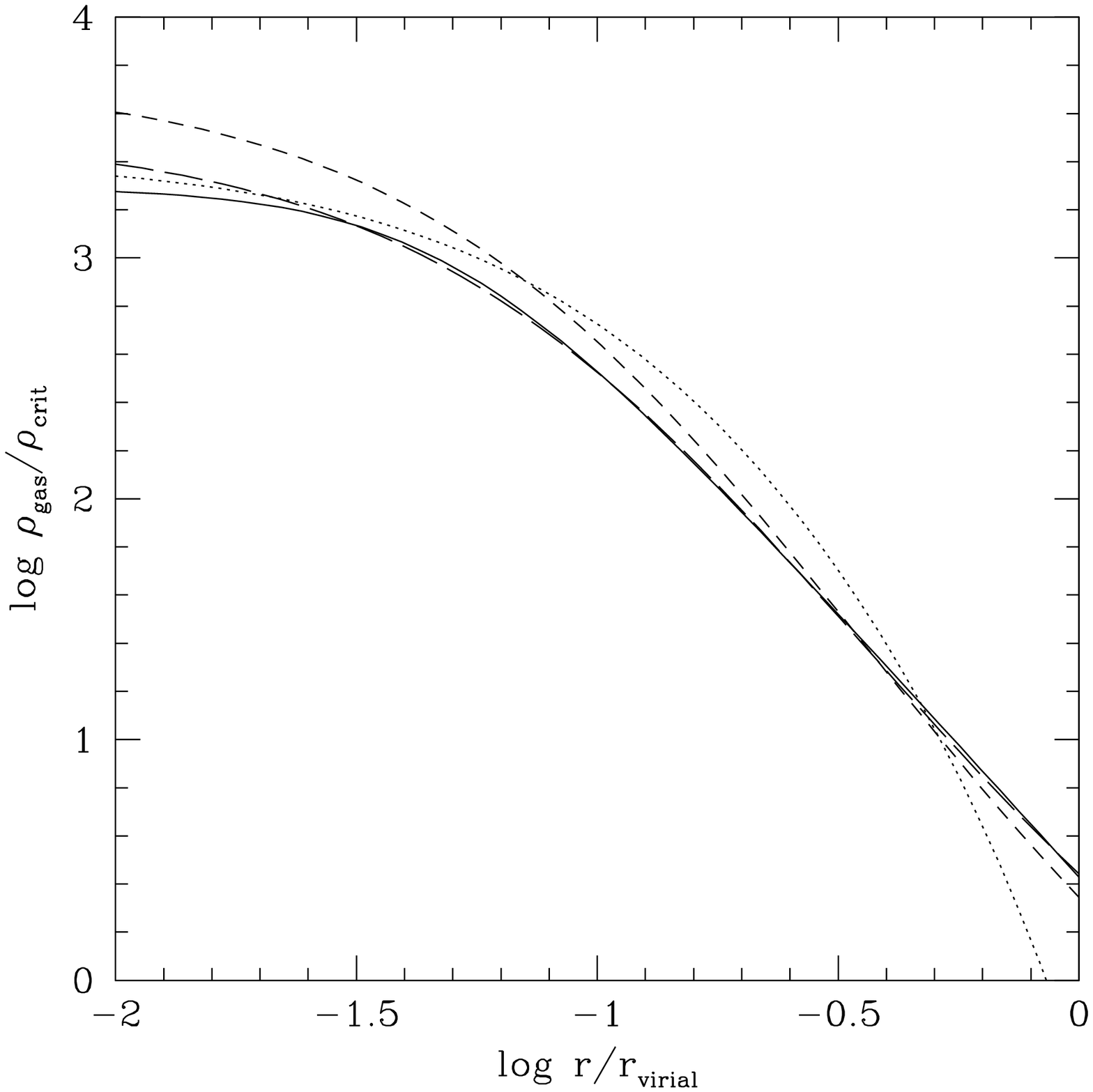}}
\caption[]{Comparison of the density profile in the following
hydrostatic polytropic models with simulated cluster $\beta$-model fit
(solid curve): $n=5$ and $\tau=1.1$ (dotted curve), $n=100$ and
$\tau=1$ (short-dashed curve), $n=100$ and $\tau=1.1$ (long-dashed
curve), where $n$ is the polytropic index and $\tau$ the ratio of the
emission-measure-weighted average temperature to virial temperature.}
\end{figure}

\begin{figure}
\centerline{\epsfxsize=0.8\textwidth\epsfbox{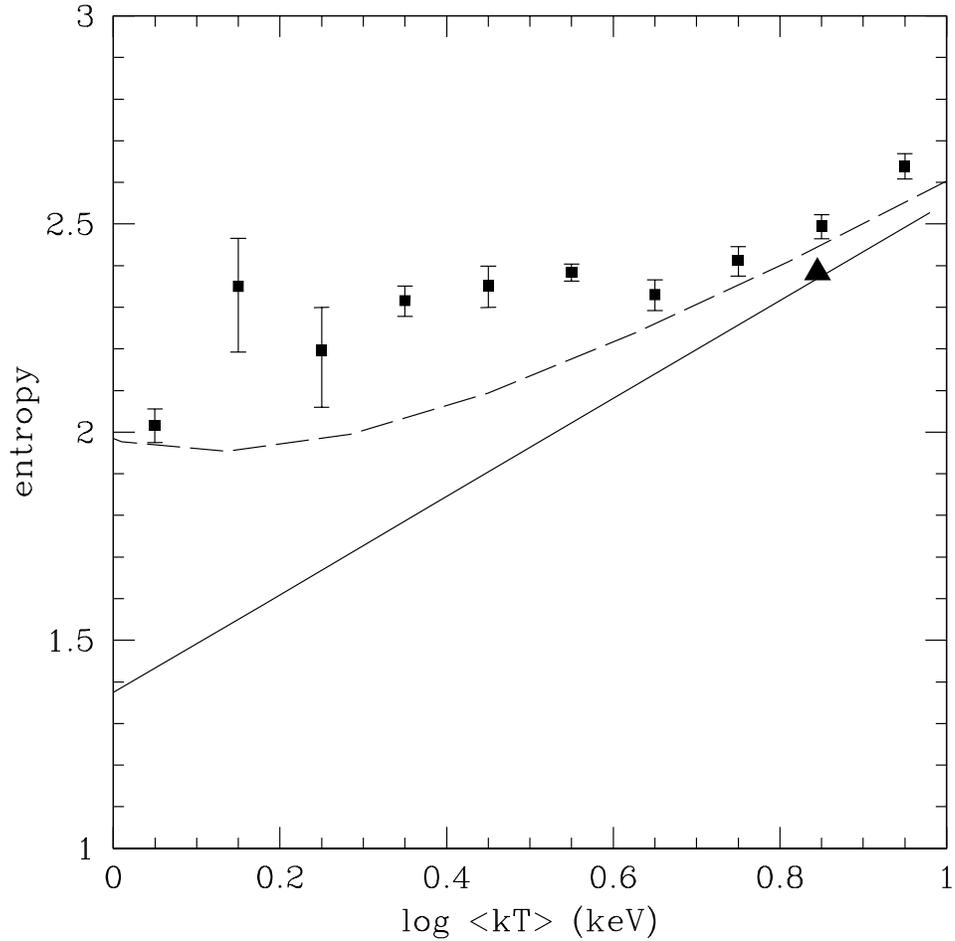}}
\caption[]{Entropy-temperature curves for heated models with $n=100$,
constant $f_{\rm virial}$, and $\epsilon=0$ (solid curve) or
$\epsilon=0.35$ (dashed curve), where the entropy is defined as ${\rm
log_{10}}(kT_x{n_e}^{-2/3})$, $n_e$ is in cm$^{-3}$ and evaluated at
$r_c$, $kT_x$ is the average (within $r_{500}$) ICM temperature in
keV, and $h_{70}=1$. The filled triangle represents the value from
numerical simulations without heating; the data points, representing
averages in ten temperature bins, and errorbars, representing the
dispersion within each bin, have been derived from {\it ASCA} data and
kindly provided by U. Hwang.}
\end{figure}

\begin{figure}
\centerline{\epsfxsize=0.8\textwidth\epsfbox{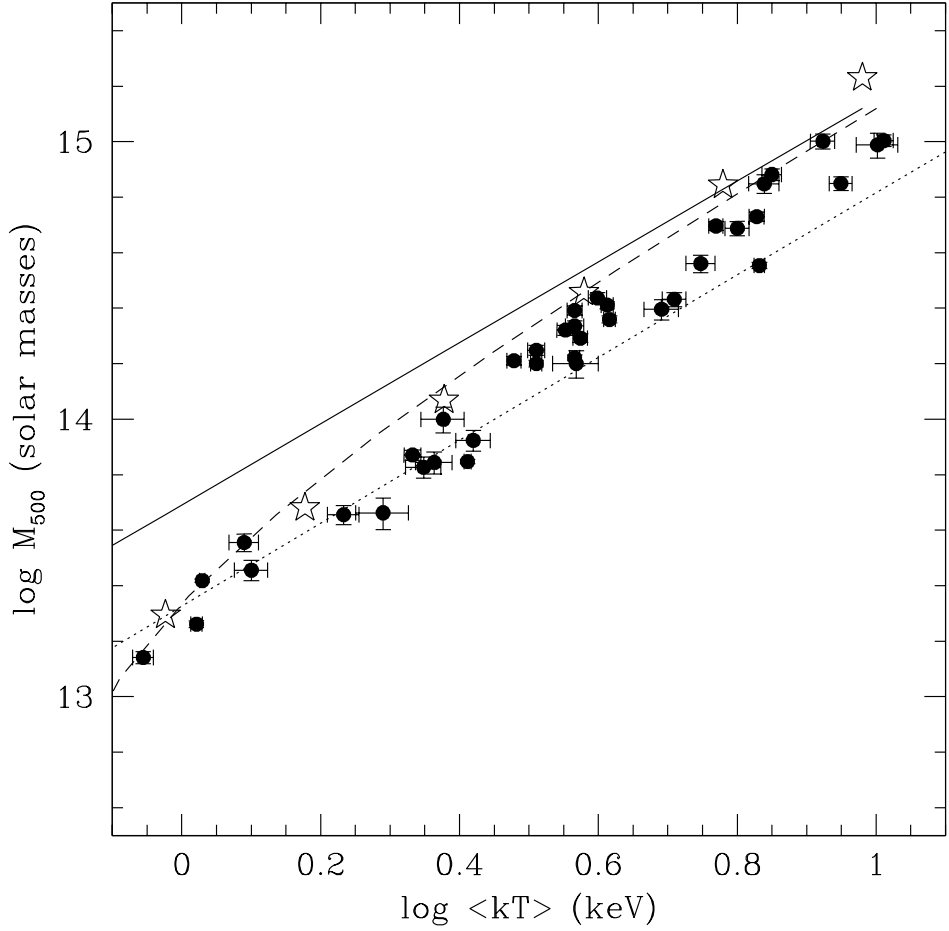}}
\caption[]{Mass-temperature curves for $\epsilon=0$ and
$\epsilon=0.35$ models as in Figure 2, best-fit observed correlation
from EF (stars), and isothermal $\beta$-model masses from Horner et
al. (1999; data points with errorbars), where the mass is the total
gravitational mass integrated out to $r_{500}$. Also shown is an
isentropic mass sequence with mass-averaged temperature approximately
equal to the virial temperature and $kT_x{n_e}^{-2/3}=200$ keV cm$^2$
(dotted curve; see Section 4.3).}
\end{figure}

\begin{figure}
\centerline{\epsfxsize=0.8\textwidth\epsfbox{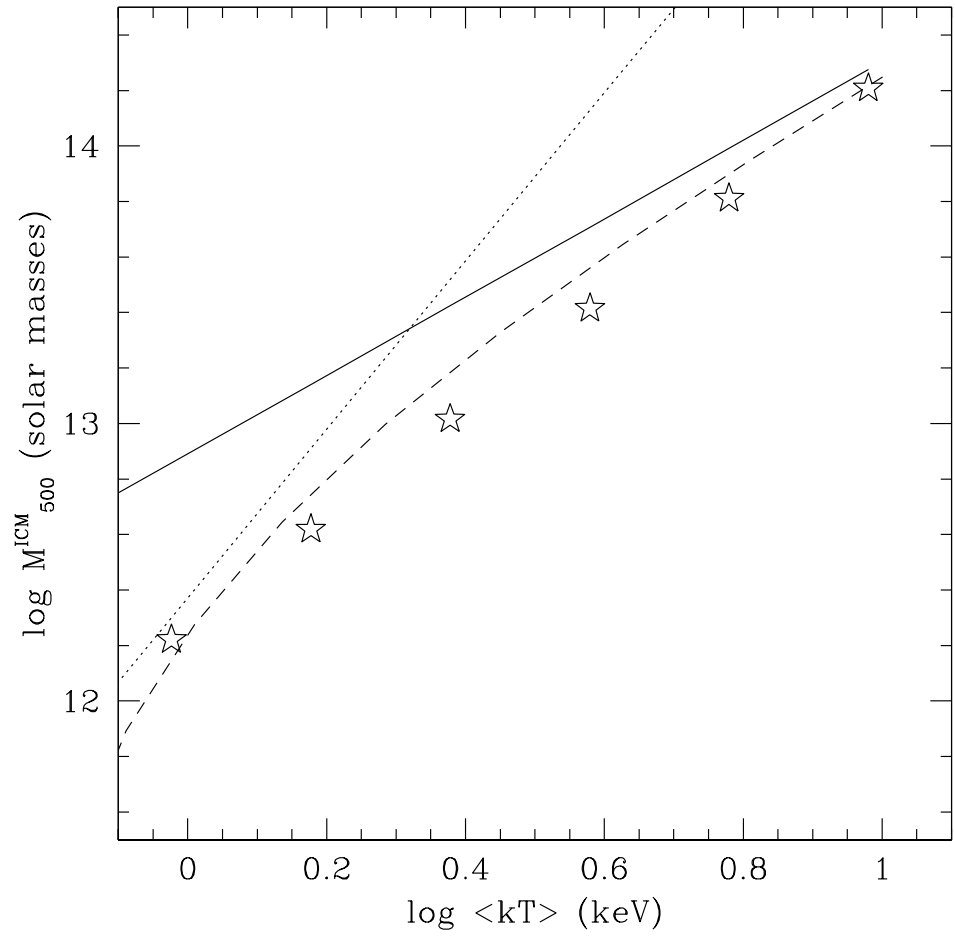}}
\caption[]{Same as Figure 3 for the ICM mass; best-fit observed
correlation is from Mohr et al. (1999).}
\end{figure}

\begin{figure}
\centerline{\epsfxsize=0.8\textwidth\epsfbox{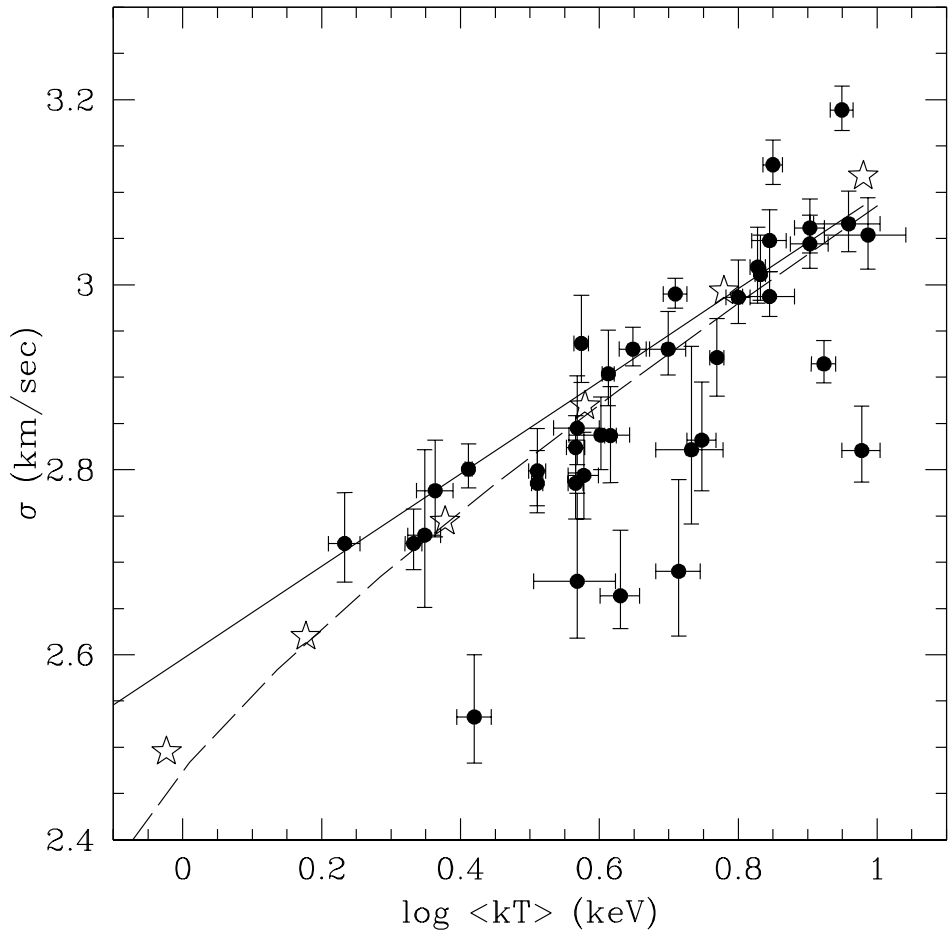}}
\caption[]{Same as Figure 3 for $(kT_{\rm virial}/\mu m_p)^{1/2}$;
observed trend is best-fit $\sigma$-$T_x$ correlation from Bird et
al. (1995); data points with errorbars are from Girardi et al.
(1998).}
\end{figure}

\begin{figure}
\centerline{\epsfxsize=0.8\textwidth\epsfbox{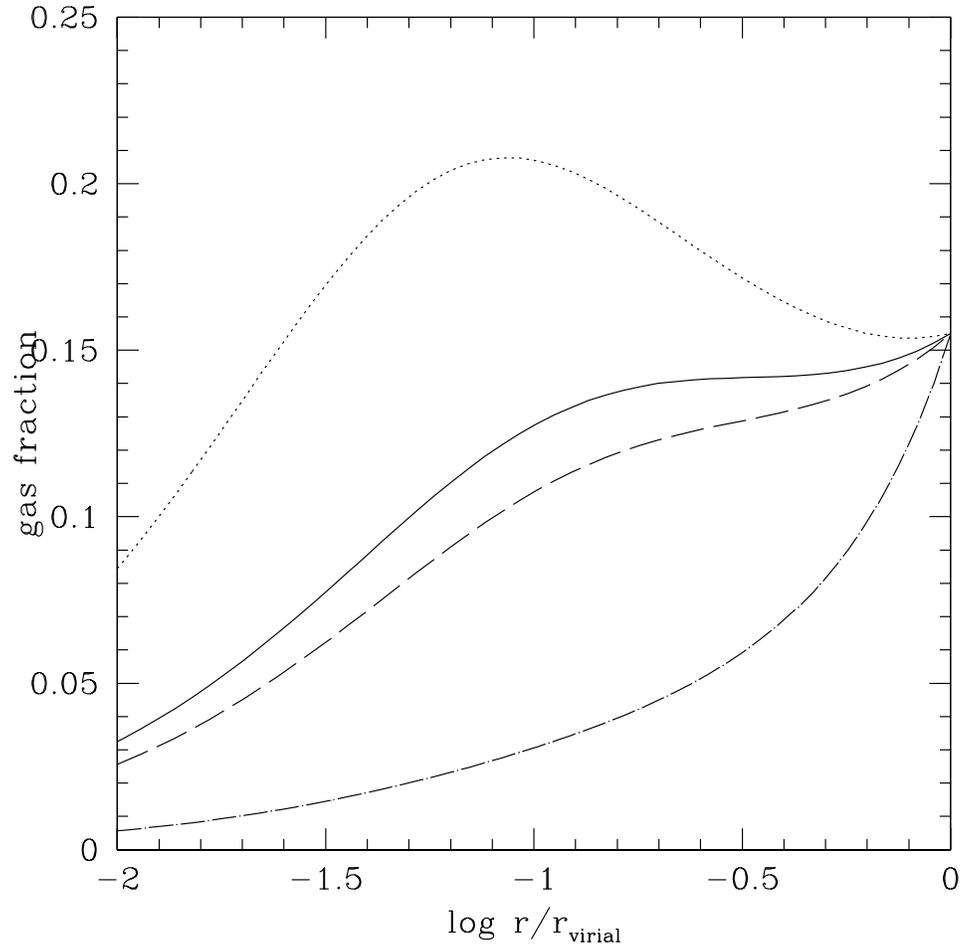}}
\caption[]{Gas fraction profiles for heated models with $n=100$,
constant $f_{\rm virial}$, and $\epsilon=0$ (solid curve: 10 keV
cluster, dotted curve: 1 keV cluster) or $\epsilon=0.35$ (dashed
curve: 10 keV cluster, dot-dashed curve: 1 keV cluster).}
\end{figure}

\begin{figure}
\centerline{\epsfxsize=0.8\textwidth\epsfbox{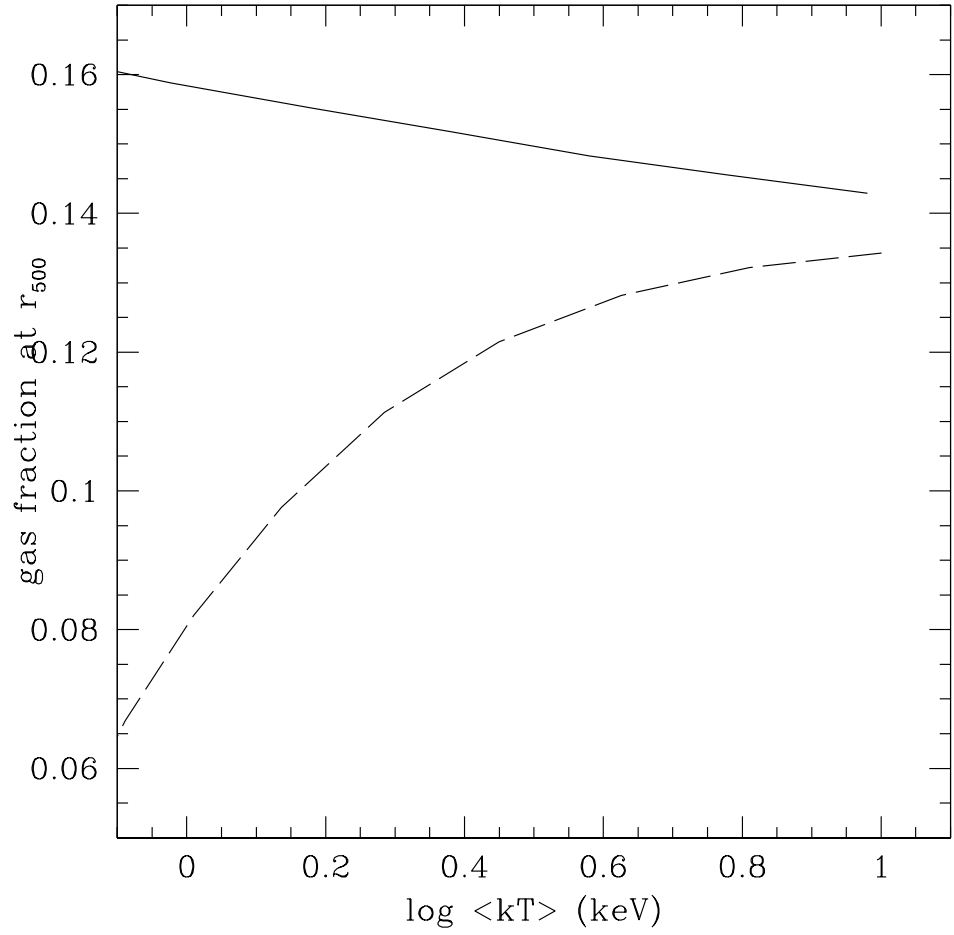}}
\caption[]{Gas fraction at $r_{500}$ for heated models with $n=100$,
constant $f_{\rm virial}$, and $\epsilon=0$ (solid curve) or
$\epsilon=0.35$ (dashed curve).}
\end{figure}

\begin{figure}
\centerline{\epsfxsize=0.8\textwidth\epsfbox{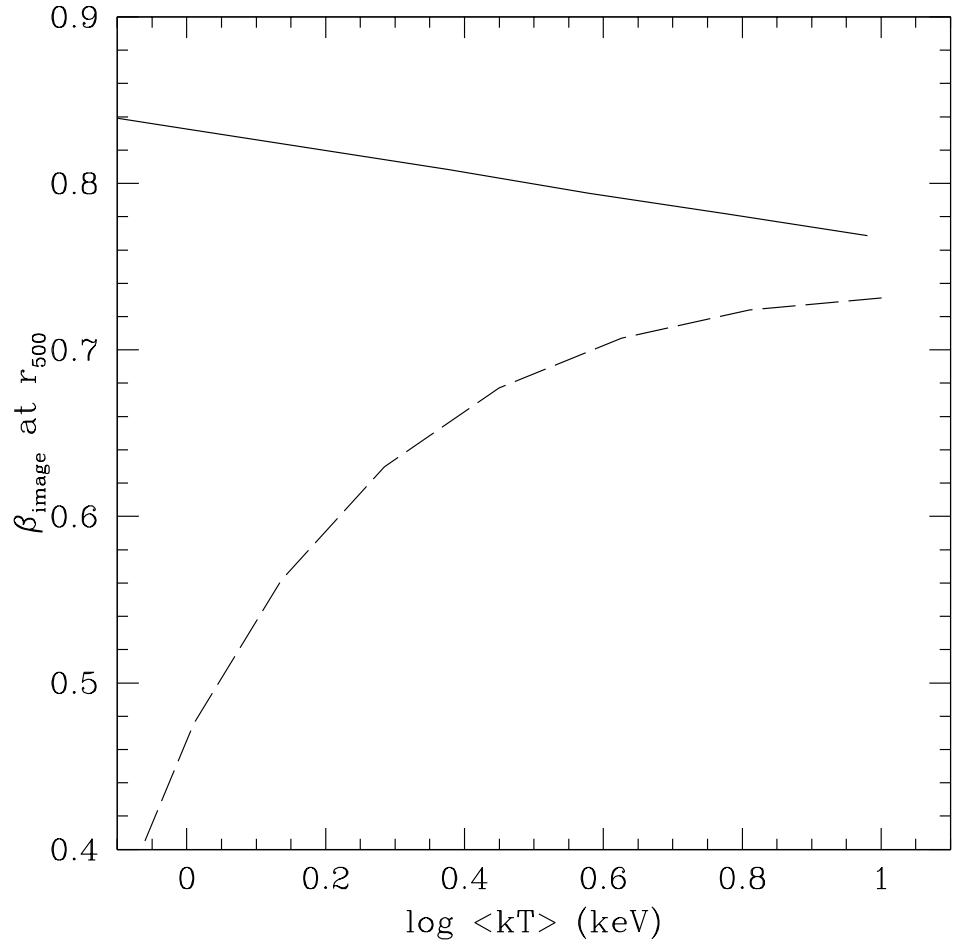}}
\caption[]{Same as Figure 7 for $\beta_{\rm image}$ (three times the
ICM density slope) at $r_{500}$.}
\end{figure}

\begin{figure}
\centerline{\epsfxsize=0.8\textwidth\epsfbox{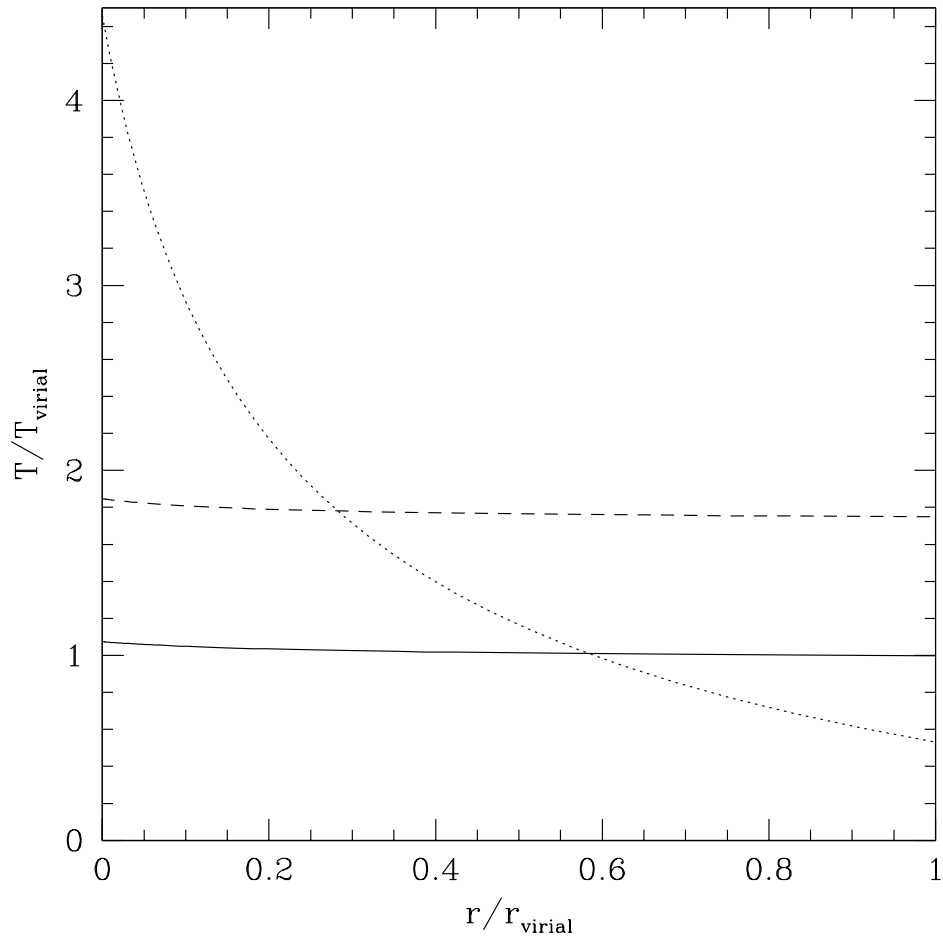}}
\caption[]{$M_{\rm virial}=2.3\ 10^{13}{\rm M}_{\odot}$ ($T_{\rm
virial}\approx 0.57$ keV) cluster temperature profiles for
$\epsilon=0$ (solid curve) and $\epsilon=0.35$ (dashed curve) heated
models featured in Figures 2-8, as well for the $kT_x{n_e}^{-2/3}=200$
keV cm$^2$ isentropic model (dotted curve). The latter two models have
emission-averaged temperatures $\approx 1$ keV, although the
mass-averaged temperature in the isentropic model is, by construction,
equal to $1.05T_{\rm virial}$.}
\end{figure}

\end{document}